\documentstyle[12pt]{article}

\begin{document}

\begin{center}

{\Large{\bf Self-similar approach to market analysis} \\ [5mm]
V.I. Yukalov} \\ [3mm]

{\it Bogolubov Laboratory of Theoretical Physics \\
Joint Institute for Nuclear Research, Dubna 141980, Russia}

\end{center}

\vskip 3cm

{\parindent=0pt

{\bf Abstract}. A novel approach to analyzing time series
generated by complex systems, such as markets, is presented. The
basic idea of the approach is the {\it Law of Self-Similar
Evolution}, according to which any complex system develops
self-similarly. There always exist some internal laws governing
the evolution of a system, say of a market, so that each of such
systems possesses its own character regulating its behaviour. The
problem is how to discover these hidden internal laws defining the
system character. This problem can be solved by employing the {\it
Self-Similar Approximation Theory}, which supplies the
mathematical foundation for the Law of Self-Similar Evolution. In
this report, the theoretical basis of the new approach to
analyzing time series is formulated, with an accurate explanation
of its principal points.}

\vskip 3cm

processes, and statistics

\newpage

{\bf 1.} {\large{\bf Law of self-similar evolution}}

\vskip 3mm

There are two sides of the approach I am going to present here:
{\it philosophical} and {\it mathematical}. The first side gives
the general idea of {\it why} should we think in these or those
terms, and the second side is supposed to answer {\it how}
concretely could we realize the idea. It is natural to start with
the formulation of the general idea.

The idea to be formulated is general since it may concern any
complex system, for instance, a market, a firm, a society, a
nation, species, a person, a brain, a heart, and so on. Shortly
speaking, a {\it complex system} is such that cannot, at least at
the present time, be completely described by any finite number of
given equations. A complex system exists as an empirical organism,
with its specific character and behaviour.

To understand the character of a system means to notice some rules
governing its behaviour and, therefore, to be able to predict the
latter. If the system is complex, such rules cannot be described
by fixed equations. But one may always characterize the behaviour
in  terms of self-similarity, implying that there exist some
specific features that, to some extent, are repeated during the
evolution of the considered system. Thus, characterizing a person,
we name the corresponding features that give us impression of what
can be expected from him or her.

The evolution of each complex system is always governed by some
laws. This is why there is a similarity in the behavioral facts of
this system. Such a behavioural self-similarity is easily
noticeable in the life of particular persons as well as in the
history of societies, nations, and biological species. This
observation suggests that there exists the {\it Law of
Self-Similar Evolution}, according to which the main features of
a complex system are preserved in the course of its evolution. And
if one has grasped these features, he or she should have the
feeling of what could be expected in future.

When the behavioral facts, related to the evolution of a complex
system, can be associated with some quantities, one gets time
series. These can be market prices or indices, measurable signals
from a brain or from a heart, some data characterizing societies
or species etc. The examples of time series are numerous, their
consideration being usually based on statistical analysis and the
construction of dynamic regression models [1--5]. These models
provide a rather reasonable description of sufficiently stable
dynamics with repeated events, like seasonal variations, but they
fail in treating such unstable systems as markets. Financial
markets are very difficult to predict. Perhaps the most central
question in finance  is under what circumstances is prediction
possible at all?

The widespread opinion is that market prices fluctuate absolutely
randomly and, thus, are unforecastable. This point of view is what
is called the Efficient Markets Hypothesis, which can be traced
back to Samuelson [6]. According to this hypothesis, in an
informationally efficient market, properly anticipated prices
fluctuate randomly. This includes the assumption that prices fully
incorporate the expectations and information of all market
participants, that all market traders are identical in the sense
that all of them possess the same and the whole information on a
market, all traders having the same physical and mental abilities.
Such a hypothesis that all investors are fully rational agents
that instantaneously and correctly process all available
information is clearly unrealistic. Moreover, people, thanks God,
are never identical neither in their abilities nor in their
wishes. They are not elementary quantum particles, but each of
them is a very complex system by its own. There is also increasing empiric
evidence that even the most competitive markets are not strictly
efficient [7--9].

A word of caution is called for with  respect to the meaning of
the term "random", which is often confused with "chaotic". These
two notions are rather different. A realistic process can be
chaotic, at the same time, possessing properties of both
regularity and randomness in different proportions, so that it is,
in principle, possible to distinguish chaos from randomness, and
even control or erase chaos and make predictions [10,11].

Dynamics of all realistic complex systems always exhibits some
part of randomness, either due to internal reasons, specific for
nonlinear dynamical systems, or caused by external stochastic
noise. Therefore randomness is inavoidable to this or that extent
and, for some time intervals, it can mask the existence of
underlying tendencies and persisting trends. Nevertheless, such
trends are to be noticeable on average, whether this concerns the
evolution of financial markets or biological species. Persisting
features retained during this evolution imply that the complex
system evolves self-similarly. It is worth emphasizing that the
Law of Self-Similar Evolution does not mean and in no sense
requires that there should be displayed some rigidly  fixed
properties of the considered system, but it rather tells that
there must exist some trends in evolution. The {\it evolutional
self-similarity} is not a static notion requiring the occurrence
of stationary fixed properties but it is a {\it dynamic concept}
stating the persistence of underlying trends.

In this section, the Law of Self-Similar Evolution has been
described as a philosophical category, in rather general and
perhaps vague words. The corresponding mathematical realization
will be presented in the following sections. The suggested
approach is mathematically based on the Self-Similar Approximation
Theory whose application to the analysis of asymptotic series is
expounded in Sec. 2. The self-similar extrapolation of asymptotic
series can be directly reformulated as a self-similar forecasting
for time series, which is explained in Sec. 3. A method for
evaluating the probabilities of self-similar patterns is developed
in Sec. 4. These sections present, for the first time, the
complete mathematical foundation of the self-similar approach to
analyzing arbitrary time series. Specifications related to market
time series are also discussed. Principal points of the approach
are summarized in Sec. 5.

\vskip 5mm

{\bf 2.} {\large{\bf Self-similarity in asymptotic series}}

\vskip 3mm

Assume that we are interested in finding a function $f(x)$ of a
real variable $x$. Let this function be defined by so complicated
equations that we are able to extract from them only asymptotic
expansions in the vicinity of some point $x_0$. Without the loss
of generality, the expansion point may be taken as zero, $x_0=0$.
Let us have several such expansions,
\begin{equation}
\label{1}
f(x) \simeq f_k(x) \qquad (x\rightarrow 0) \; ,
\end{equation}
enumerated by the index $k=0,1,2,\ldots$. The basic problem is what
can be said about the value of $f(x)$ at finite $x$ if all we know
are asymptotic expansions $f_k(x)$ in the vicinity of
$x\rightarrow 0$? This problem is constantly met in physics and
applied mathematics, where it is often called the problem of
function reconstruction or the problem of summation of asymptotic
series [12].

An original general approach, named the {\it Self-Similar
Approximation Theory}, for reconstructing functions from a set of
their approximate expressions has been developed and successfully
applied to various problems [13--23]. The name comes from the
basic idea of the approach to present the passage between
subsequent approximations as a self-similar transformation. More
precisely, it was shown [15--19] that for an approximation
sequence $\{ f_k(x)\}$ it is possible to construct a cascade, that
is a dynamical system with discrete time, whose trajectory is
bijective to the sequence $\{ f_k\}$, so that the sought function
$f(x)$ corresponds to a fixed point of the cascade. If we treat
the given complicated equations, together with a calculational
algorithm, as a complex system generating a sequence $\{ f_k\}$,
then the latter is nothing but a prototype of a time series, the
approximation number $k$ playing the role of time. Hence the
self-similar approximation theory is the mathematical realization
of the law of self-similar evolution.

Asymptotic expansions are usually presented as power-law series
\begin{equation}
\label{2}
f_k(x) = \sum_{n=0}^k\; a_n\; x^{\alpha_n} \; ,
\end{equation}
where $\alpha_n$ are arbitrary real numbers arranged in ascending
order,
\begin{equation}
\label{3}
\alpha_n <\alpha_{n+1} \qquad (n=0,1,\ldots,k) \; .
\end{equation}
For the purpose of self-similar analysis, the presentation of the
series (2) has to satisfy several general properties:

\vskip 2mm

(i) Analysis should not depend on the choice of units for the
variable. Hence the latter is to be taken in a dimensionless form.

\vskip 2mm

(ii) Asymptotic expansions are to be reduced to a scale-invariant
form. To this end, it is always possible to factor out the term
\begin{equation}
\label{4}
f_0(x) = a_0\; x^{\alpha_0} \qquad (a_0\neq 0)
\end{equation}
and to introduce the scale-invariant function
\begin{equation}
\label{5}
\varphi_k(x) \equiv \frac{f_k(x)}{f_0(x)} \; ,
\end{equation}
where $x$ is assumed to be dimensionless. Evidently, function (5)
does not depend on the change of scales for $f_k$.

\vskip 2mm

(iii) If the variable $x$ pertains to a finite interval, the
latter is to be normalized to the unitary interval. So that
everywhere in what follows, it is assumed that $x\in[0,1]$.

\vskip 2mm

Note that if the variable pertains to an infinite interval, one
may pursue different ways depending on what additional information
on the behaviour of $f(x)$ at $x\rightarrow\infty$ is available.
If there is no such information, it is possible to transform the
infinite interval to the unitary one by means of the change of
variables $x'=x/(1+x),\; x=x'/(1-x')$, so that $x'\in[0,1]$.

For the scale-invariant function (5), we have
\begin{equation}
\label{6}
\varphi_k(x) = \sum_{n=0}^k\; b_n\; x^{\beta_n} \; ,
\end{equation}
where $x\in[0,1]$ and
\begin{equation}
\label{7}
b_n\equiv \frac{a_n}{a_0} \; , \qquad \beta_n \equiv
\alpha_n -\alpha_0 \geq \beta_0 = 0 \; .
\end{equation}
Clearly, $\varphi_0(x)=1$. Also, all powers $\beta_n$ are
positive, even if some $\alpha_n$ are negative, which follows from
the ascending order (3). Recall that, by construction, the series
(6) are assumed to be asymptotic, having sense only for
$x\rightarrow 0$, while for any finite $x$ sequence
$\{\varphi_k(x)\}_{k=0}^{\infty}$ diverges. It is possible to say
that the latter sequence converges just at one point $x=0$.

To proceed further, we have to transform divergent series to a
form that would have sense for finite $x\in[0,1)$. This can be
done with the help of control functions [13], which can be
introduced in different ways [13--23]. Dealing with asymptotic
series, it is convenient to invoke the multiplicative power-law
transformation [20--22] defined as
\begin{equation}
\label{8}
\Phi_k(x,s) \equiv x^s\; \varphi_k(x) \; ,
\end{equation}
with the inverse transformation
\begin{equation}
\label{9}
\varphi_k(x) = x^{-s}\;\Phi_k(x,s) \; .
\end{equation}
Since power laws are common in describing fractal objects,
equation (8) may be called the {\it fractal transformation}.
The transform (8), according to equation (6), is
\begin{equation}
\label{10}
\Phi_k(x,s) =\sum_{n=0}^k\; b_n\; x^{s+\beta_n} \; .
\end{equation}
Here $s=s(x)$ is a control function, whose role is to make the
series (10) meaningful for finite $x$. These series can be
considered as an expansion in powers of the new variable $x^s$. As
is evident, such series are asymptotic with respect to
$x^s\rightarrow 0$. The latter limit can be achieved if, instead
of forcing $x$ to zero, we keep $|x|<1$ and setting
$s\rightarrow\infty$. Thus, the series (10) can be treated as
asymptotic with respect to $s\rightarrow\infty$ for all $|x|<1$.
Now we may say that the sequence $\{\Phi_k(x,s)\}_{k=0}^\infty$
converges for all $|x|<1$, provided that $s\rightarrow\infty$.
In this way, we come to the natural choice of the control function
$s\rightarrow\infty$. Recall that we are considering the case when
no additional constraints are imposed on the behaviour of the sought
function and all we know are its asymptotic expansions (1). In the
intermediate expressions, the value of $s$ is assumed to be
asymptotically large, and the actual limit $s\rightarrow\infty$ is
to be taken after the inverse transformation (9).

Since our consideration here concerns functions, we need to define
the property of {\it functional self-similarity}, which should not
be confused with geometric self-similarity describing fractals.
The notion of geometric self-similarity [24] is connected with the
scaling of a variable, which is only a particular kind of the more
general notion of the functional or {\it group self-similarity}
[15--19]. To correctly define the latter, we need to introduce
some notation. We define the expansion function $x(\varphi,s)$ by
the equation
\begin{equation}
\label{11}
\Phi_0(x,s) =\varphi\; , \qquad x=x(\varphi,s) \; .
\end{equation}
With the form (10), this gives $x(\varphi,s)=\varphi^{1/s}$ and
$\varphi=x^s$. Introduce the mapping
\begin{equation}
\label{12}
y_k(\varphi,s)\equiv \Phi_k(x(\varphi,s),s) \; .
\end{equation}
Let $\Phi_k(x,s)$ be real for all $k=0,1,2,\ldots$ and all
$x\in[0,1]$. From $\Phi_k(x,s)\in{\bf R}$ it follows that
$y_k(\varphi,s)\in{\bf R}$ for all $k=0,1,2,\ldots$ and
$\varphi\in{\bf R}$. Therefore the mapping (12) is an endomorphism
on ${\bf R}$. It is this endomorphism that allows us to formulate
the property of group self-similarity we need.

Our aim is to present the change of the endomorphism (12), when
varying the approximation number $k$, as the evolution of $y_k$
with respect to the discrete time $k$. From the point of view of
group theory, self-similarity is nothing but a semigroup property.
The latter, for the evolution of $y_k$ with respect to $k$, reads
$y_{k+p}=y_k\cdot y_p$. As follows from the definition (11), the
unit element is $y_0$, since $y_0(\varphi,s)=\varphi$. The family
of endomorphisms, $\{ y_k|\; k=0,1,2,\ldots\}$, with the semigroup
property forms a dynamical system in discrete time, called the
cascade. The semigroup property, in terms of the notation (12),
takes the form
\begin{equation}
\label{13}
y_{k+p}(\varphi,s) =y_k(y_p(\varphi,s),s) \; .
\end{equation}
Since in the accepted interpretation, we treat the sequence
$\{ y_k(\varphi,s)\}$ as a trajectory resulting from the evolution
of the cascade $\{ y_k|\; k=0,1,2,\ldots\}$, the relation (13) may
be called the {\it evolutional self-similarity}. As far as the
corresponding semigroup property is natural for dynamical systems,
the equation (13) may also be termed the {\it dynamic
self-similarity}. This equation (13) is a necessary condition for
the fastest-convergence criterion [16,17]; the cascade fixed point
representing the sought function.

Following the general theory [14--19], the cascade $\{ y_k|\;
k=0,1,2,\ldots\}$ can be embedded into a flow $\{ y_\tau|\;
\tau\geq 0\}$, which is a dynamical system in continuous time. For
the latter, one may write the Lie equation which is a differential
equation of motion. The flow velocity field is defined, by means
of the Euler discretization, as the cascade velocity
\begin{equation}
\label{14}
v_n(\varphi,s) \equiv y_n(\varphi,s) - y_{n-1}(\varphi,s) \qquad
(n=1,2,\ldots,k) \; .
\end{equation}
From equations (10) - (12), it follows that
\begin{equation}
\label{15}
y_k(\varphi,s) =\sum_{n=0}^k \; b_n\; \varphi^{1+\beta_n/s} \; ,
\end{equation}
which results in
\begin{equation}
\label{16}
v_n(\varphi,s) = b_n\; \varphi^{1+\beta_n/s} \; .
\end{equation}
The differential equation of motion can be integrated. The
integration over the effective time goes from $\tau=n$ to
$\tau=n+\tau_n$, with $\tau_n$ being the effective time required
for reaching a fixed point after the $n$-th step. In this way,
\begin{equation}
\label{17}
\int_n^{n+\tau_n} \; dt = \tau_n \; .
\end{equation}
Therefore, the {\it evolution integral} [15--19] acquires the form
\begin{equation}
\label{18}
\int_{y_n}^{y_n^*}\; \frac{d\varphi}{v_n(\varphi,s)}= \tau_n\; .
\end{equation}
This, with the cascade velocity (16), yields
$$
\Phi^*_n =\left ( \Phi_{n-1}^{-\beta_n/s} - \;
\frac{\beta_n}{s} \; b_n \tau_n\right )^{-s/\beta_n} \; ,
$$
where $\Phi_n^*=\Phi_n^*(x,s)\equiv y_n^*(x^s,s)$ and
$\Phi_n=\Phi_n(x,s)$. In particular,
$$
\Phi^*_1(x,s) = x^s \left ( 1  - \; \frac{\beta_1}{s} \;
b_1\tau_1\; x^{\beta_1} \right )^{-s/\beta_1} \; .
$$

Returning back to the function $\varphi_k^*(x)$ by means of the
inverse transformation (9), we have to take the limit
$s\rightarrow\infty$. For example,
$$
\varphi_1^*(x) \equiv \lim_{s\rightarrow\infty}\; x^{-s}\;
\Phi_1^*(x,s) \; ,
$$
which results in the first-order self-similar approximant
$$
\varphi_1^*(x) =\exp\left ( b_1\tau_1\; x^{\beta_1}\right ) \; .
$$
To obtain higher approximations, we can accomplish such a
renormalization procedure $2k$ times for each $\Phi_k^*(x,s)$,
which was called [21,22] {\it self-similar bootstrap}. However the
same result can be reached twice faster, by accomplishing $k$
renormalizations, in the following way. We may present the
function (6) as
$$
\varphi_k(x) = 1 + b_1\; x^{\beta_1}\left ( 1 +
\frac{b_2}{b_1}\; x^{\beta_2-\beta_1}\left ( 1 +
\frac{b_3}{b_2}\; x^{\beta_3-\beta_2} \left ( 1 +\ldots \right )
\right ) \ldots\right ) \; .
$$
The latter can be written in the form $\varphi_k(x)=1+x_1$, in which
$x_1$ is expressed through $x_2$, and $x_2$ through $x_3$, and so on
according to the rule
$$
x_n =\frac{b_n}{b_{n-1}}\; x^{\beta_n-\beta_{n-1}}\left ( 1 +
x_{n+1} \right ) \; ,
$$
where $n=1,2,\ldots,k$. Considering each $x_n$ as a small
parameter, we need to accomplish $k$ times the first-order
renormalization procedure described above. As a result,
introducing the notation
\begin{equation}
\label{19}
c_n \equiv \frac{a_n}{a_{n-1}}\; \tau_n \; , \qquad \nu_n\equiv
\alpha_n -\alpha_{n-1} \qquad (n=1,2,\ldots,k) \; ,
\end{equation}
we come to the $k$-order {\it self-similar exponential
approximant}
\begin{equation}
\label{20}
\varphi_k^*(x)= \exp\left ( c_1x^{\nu_1}\; \exp\left ( c_2x^{\nu_2}
\ldots\exp\left ( c_kx^{\nu_k}\right ) \right ) \ldots\right )\; .
\end{equation}
This, for short, can also be named the $k$-order {\it
superexponential}.

It is worth emphasizing that although the form (20) reminds the
Euler nested exponentials [25,26], it is principally different
from the latter. First of all, the Euler superexponentials are
defined only for integer powers $\alpha_n$. Second, when one tries
to sum power series by means of such continued exponentials, one
fits the coefficients in the latter so that to reproduce those in
the power series, as a result of which the constructed
superexponentials have the same radius of convergence as the
related power series on the real axis [27,28].

In our case, the self-similar exponential (20) contains the
coefficients $c_n$ given by equation (19), where the effective
renormalization time $\tau_n$ is yet undefined. The latter plays
the role of a control function that is to be determined from
additional conditions. One way could be to find $\tau_n$ from the
fixed-point conditions [22] having the form of the
minimal-difference criterion [13]. Such fixed-point equations not
always possess solutions. Here we suggest another, and more
general way of defining the control functions $\tau_n$. This
method is commonly employed for defining control functions in the
optimal control theory. The idea is to construct a cost functional
whose minimization yields the control functions of interest.

The quantity $\tau_n$ appears in the evolution integral (18),
where it has the meaning of an effective time required for reaching,
after the $n$-th step, a fixed point representing the sought
function. One, clearly, would like to reach the answer as fast as
possible. The minimal number of renormalization steps is,
obviously, one. Therefore one would like that the total effective
time $n\tau_n$ be also close to one. Hence, one should look for
$\tau_n$ being close to $1/n$. At the same time, how fast one
reaches the fixed point depends on the distance of the latter from
the starting point. The distance that is passed during the time
$\tau_n$, with a velocity $v_n$, can be evaluated as $v_n\tau_n$.
Thus, we need to find a minimal time $\tau_n$, being close to
$1/n$, corresponding to the minimal distance $v_n\tau_n$.
This suggests us to construct the {\it time-distance cost
functional}
\begin{equation}
\label{21}
F =\frac{1}{2}\; \sum_n \left [ \left ( \tau_n -\;
\frac{1}{n}\right )^2 + \left ( v_n\tau_n\right )^2\right ]\; ,
\end{equation}
whose minimization with respect to the control function $\tau_n$
yields
\begin{equation}
\label{22}
\tau_n =\frac{1}{n(1+v_n^2)}\; .
\end{equation}
The velocity $v_n=v_n(x)$ is to be understood as the image, in the
domain of $x$, of the cascade velocity (16), with taking account
of the inverse transformation (9), which gives $v_n(x) \equiv
x^{-s}v_n(x^s,s)$. From here, one has
\begin{equation}
\label{23}
v_n(x) =\varphi_n(x) -\varphi_{n-1}(x) = b_n\; x^{\beta_n} \; ,
\end{equation}
which defines the control function $\tau_n=\tau_n(x)$ according to
the expression (22).

In this way, we find the {\it controllers}
$$
c_n(x) =\frac{a_n}{a_{n-1}}\; \tau_n(x) \; , \qquad \nu_n=
\alpha_n -\alpha_{n-1} \; ,
$$
\begin{equation}
\label{24}
\tau_n(x) =\frac{1}{n[1+v_n^2(x)]} \; , \qquad
v_n(x) =\frac{a_n}{a_0}\; x^{\alpha_n-\alpha_0} \; .
\end{equation}
Combining equations (5) and (20), we obtain the $k$-order
self-similar exponential approximant
\begin{equation}
\label{25}
f_k^*(x) = f_0(x) \exp\left ( c_1 x^{\nu_1}\; \exp\left (
c_2x^{\nu_2}\ldots \exp\left ( c_kx^{\nu_k}\right )\right )
\ldots\right ) \; ,
\end{equation}
in which $c_n=c_n(x)$, and $f_0(x)$ is given by formula (4). The
approximant (25) extrapolates the asymptotic series (2), valid only
for $x\rightarrow 0$, to the region of finite $x\in[0,1)$. The
value of $f(x)$ at the point $x=1$ can be defined as the limit from
the left, as $x\rightarrow 1-0$. Therefore the superexponential
(25) extrapolates the sought function $f(x)$ from asymptotically
small $x\rightarrow 0$ to the whole unitary interval $0\leq x\leq 1$.

It is worth recalling that in the asymptotic series (2) the powers
$\alpha_n$ were assumed to be arbitrary real numbers, with the
sole requirement that they are arranged in the ascending order
(3). Some, or even all, of these powers could be negative. If so,
then the initial term (4), with $\alpha_0<0$, has the power-law form
that has been so much discussed in literature. The asymptotic
existence of power laws is well known in critical phenomena. The
relevance of power laws has repeatedly been claimed to describe
many natural phenomena, ranging from earthquakes [24,29,30] to
different economic and financial distributions [31,32]. Since such
power laws are practically always asymptotic, their more general
form should include corrections leading to the power-law series
(2). Extrapolating these asymptotic series in the described
self-similar way, one should come to the self-similar
exponentials. The first-order approximation then results in a kind
of a stretched exponential. The stretched exponential
distributions describe many phenomena in nature and economy either
not worse or even better than power-law distribution functions
[33]. More generally, the extrapolation of power laws should lead
to the self-similar nested exponentials (25), whose structure
evidently demonstrates the existence of many scales.

\vskip 5mm

{\large{\bf 3. Self-similarity in time series}}

\vskip 3mm

The technique of the self-similar extrapolation for asymptotic
series can be reformulated as the method of forecasting for time
series. It is necessary to call attention to the existence of
several principal points in this reformulation. Overlooking these
points would essentially restrict the applicability of the method.
The correct general way of self-similar forecasting for time
series is advanced below.

First of all, the same basic requirements that were imposed on
asymptotic series are compulsory for time series:

\vskip 2mm

(i) The measured quantity is to be presented in a scale-invariant
form. This can be easily done by normalizing the given data to the
value $f_0$ of the measured quantity at the initial time, which is
analogous to introducing the scale-invariant function (5).

\vskip 2mm

(ii) The time variable has to be normalized to a dimensionless
form, such that the prediction time would pertain to the unitary
interval $[0,1]$. This normalization eliminates the ambiguity in
defining the power of the power-law transformation (8), requiring
that $s\rightarrow 0$, which results in self-similar
superexponentials (20).

In addition to these requirements, common both for asymptotic as
well as for time series, there arises a question specific for time
series: how the latter should be presented, as a backward or
forward recursion? The answer to this question follows from the
comparison of asymptotic series with time series. In order that
the extrapolation of asymptotic series could be directly extended
to forecasting for time series, the prediction horizon is to be a
unitary interval and the available information from the past has
to provide approximate forecasts for asymptotically small time
$t\rightarrow +0$. This requirement can be strictly accomplished
only for backward recursion. Hence, we have one more restriction,
specific for time series:

\vskip 2mm

(iii) For the correct usage of the self-similar analysis, time
series are to be arranged as a {\it backward recursion}. This
means that, if we are given a set $\{ f_n\}$ of data $f_n$
corresponding to a quantity of interest, measured at the times
$t_n$, where $n=0,1,2,\ldots,k$, the moments of time are to be
ordered so that
\begin{equation}
\label{26}
t_{n+1} < t_n \qquad (n=0,1,2,\ldots,k) \; ,
\end{equation}
with the initial time $t_0=0$. The past-history {\it data base} of
$k$-th order is the set
\begin{equation}
\label{27}
{\bf D}_k =\left\{ f_k,f_{k-1},\ldots,f_0|\; t_k<t_{k-1} <
\ldots <0 \right\} \; .
\end{equation}

To start the procedure of self-similar forecasting for future
times $t\in[0,1]$, we need to possess a sequence of approximations
valid for asymptotically small $t\rightarrow +0$. The role of such
asymptotic forecasts can naturally be played by functions $f_k(t)$
interpolating the given data base (27) for the past time horizon
$t_k\leq t\leq 0$, so that
\begin{equation}
\label{28}
f_k(t_n) = f_n \qquad (n=0,1,\ldots,k) \; .
\end{equation}
This interpolation can be uniquely defined by the Lagrange
interpolation formula [34] presenting the interpolation function
\begin{equation}
\label{29}
f_k(t) =\sum_{n=0}^k \; f_n\; l_n^k(t) \qquad (k\geq 1)
\end{equation}
as a series over the Lagrange polynomials
\begin{equation}
\label{30}
l_n^k(t) \equiv \prod_{m(\neq n)}^k
\frac{t-t_m}{t_n-t_m} \qquad (n\leq k) \; .
\end{equation}
Because of the property $l_n^k(t_m)=\delta_{mn}$ of the Lagrange
polynomials, condition (28) is automatically satisfied. Note that
the interpolation form (29) is evidently scale invariant with
respect to time and the ratio $f_k(t)/f_0$ is scale invariant with
respect to units of $f_k$. Also, one may remark that the time
moments $t_n$ are not necessarily equidistant, but can be chosen
arbitrarily.

The interpolation formula (29) can be rewritten as the algebraic
polynomial
\begin{equation}
\label{31}
f_k(t) = \sum_{n=0}^k\; a_n\; t^n \qquad (a_0=f_0) \; ,
\end{equation}
with the coefficients $a_n\equiv a_{nk}$ immediately following
from equations (29) and (30). The values of these coefficients
depend, of course, on the given data base (27), but for the
simplicity of notation, we shall omit in what follows the
additional index $k$. Consider the subsequence $\{
f_n(t)\}_{n=0}^k$ of the terms
\begin{equation}
\label{32}
f_n(t) \equiv \sum_{m=0}^n \; a_m\; t^m \; ,
\end{equation}
which, for a given $k$, tends to the polynomial (31) as
$n\rightarrow k$ and for which
\begin{equation}
\label{33}
f_n(0) = a_0 = f_0 \qquad (n=0,1,\ldots,k) \; .
\end{equation}
Then the subsequence $\{ f_n(t)\}_{n=0}^k$ can be treated as a
sequence of approximations asymptotically valid for
$t\rightarrow +0$. Employing the self-similar extrapolation,
described in the previous section, we may construct a forecast for
the time interval $0\leq t\leq 1$. The result is given by formulas
(24) and (25), where we need to change the variable $x$ by the
time $t$ and to put $\alpha_n=n$, $\nu_n=1$. Instead of equations
(24), we have the {\it controllers}
$$
c_n(t) =\frac{a_n}{a_{n-1}} \; \tau_n(t) \qquad (n=1,2,\ldots,k)
$$
\begin{equation}
\label{34}
\tau_n =\frac{1}{n[1+v_n^2(t)]} \; , \qquad v_n(t) =
\frac{a_n}{f_0}\; t^n \; .
\end{equation}
And the self-similar exponential (25) takes the form
\begin{equation}
\label{35}
f_k^*(t) = f_0\exp\left ( c_1 t\;\exp\left ( c_2 t \ldots\exp
\left( c_k t\right ) \right ) \ldots \right ) \; ,
\end{equation}
in which $c_n=c_n(t)$. Expression (35) is the {\it self-similar
forecast} for the future time interval $0 \leq t\leq 1$, predicted
on the grounds of the past-history data base (27).

\vskip 5mm

{\large{\bf 4. Ensemble of possible scenarios}}

\vskip 3mm
For each given data base (27), one may construct the self-similar
forecast (35). But one can take several different data bases by
varying either the data-base order $k$, or by changing the time
intervals between the moments of time $t_n$, where
$n=0,1,\ldots,k$, or by varying both the data-base order as well
as the data-base scale. So that, in general, one can consider an
ensemble of different data bases. Each of the latter has to be
labelled by two indices,
\begin{equation}
\label{36}
{\bf D}_k(j) =\left\{ f_k^{(j)},f_{k-1}^{(j)},\ldots,f_0|\;
t_k^{(j)}< t_{k-1}^{(j)}< \ldots < 0 \right\} \; ,
\end{equation}
one index, $k$, defining the data base order, and another, $j$,
specifying the chosen time scale of the past. For each data base
(36), one obtains a forecast $f_k^*(j,t)$ according to the rule
(35), but with different values of $c_n=c_{nk}$. Hence there
exists an ensemble $\{ f_k^*(j,t)\}$ of possible forecasts, or
admissible scenarios. Which of these possible forecasts should one
trust?

Life teaches us that in the majority of cases nothing can be
trusted for hundred percent. But, when there can happen several
different events, they can be classified by estimating their
probabilities. Hence we need to define a probability measure on
the ensemble of scenarios $\{ f_k^*(j,t)\}$.

This problem is analogous to the problem of pattern selection
occurring for nonlinear differential equations in partial
derivatives. Such equations sometimes possess a set of solutions
corresponding to different spatio-temporal structures, or patterns
[35]. A general approach for treating the problem of pattern
selection has been suggested [36]. This approach
can be directly applied for weighting possible scenarios from the
given ensemble of self-similar forecasts. For this purpose, we may
consider the passage from $f_k^*(j,t)$ to $f_{k+1}^*(j,t)$ as the
motion with respect to $k$. Then the map $\{ f_k^*(j,t)|\;
k=1,2\ldots\}$ is to be treated as the image of a dynamical system
with discrete time $k$. The probability of a scenario $f_k^*(j,t)$
can be defined as
\begin{equation}
\label{37}
p_k(j,t) =\frac{1}{Z_k(t)}\; \exp\left\{ -\Delta
S_k(j,t)\right\}\; ,
\end{equation}
where $Z_k(t)$ is a normalization factor, being the sum
$$
Z_k(t) \equiv \sum_j \exp\left\{
-\Delta S_k(j,t)\right\}
$$
over the pattern indices $j$, and the entropy variation
\begin{equation}
\label{38}
\Delta S_k(j,t) \equiv S_k(j,t) - S_1(j,t)
\end{equation}
shows the change of entropy with respect to the effective time
$k$. The entropy of a dynamical system may be defined, by analogy
with statistical systems, as the logarithm of an elementary phase
volume [36], the latter, in our case, being $|\delta f_k^*(j,t)|$.
Thus the dynamical entropy is
\begin{equation}
\label{39}
S_k(j,t) \equiv \ln|\delta f_k^*(j,t)|\; .
\end{equation}
Then the entropy variation (38) becomes
\begin{equation}
\label{40}
\Delta S_k(j,t) =\ln\left |
\frac{\delta f_k^*(j,t)}{\delta f_1^*(j,t)}\right | \; .
\end{equation}
With the notation for the mapping multiplier
\begin{equation}
\label{41}
m_k(j,t) \equiv \frac{\delta f_k^*(j,t)}{\delta f_1^*(j,t)} =
\frac{\partial f_k^*(j,t)/\partial t}{\partial f_1^*(j,t)/
\partial t} \; ,
\end{equation}
the entropy variation (40) reduces to
\begin{equation}
\label{42}
\Delta S_k(j,t) =\ln | m_k(j,t)| \; .
\end{equation}
It is convenient to introduce the average multiplier $\overline
m_k(t)$ by the relation
\begin{equation}
\label{43}
\frac{1}{|\overline m_k(t)|} \equiv \sum_j \;
\frac{1}{|m_k(j,t)|} \; .
\end{equation}
Using equations (42) and (43), for the scenario probability (37),
we have
\begin{equation}
\label{44}
p_k(j,t) = \left |
\frac{\overline m_k(t)}{m_k(j,t)} \right |\; .
\end{equation}

Note that the {\it scenario probability} (44) is normalized with respect
to the summation over the pattern indices $j$ corresponding to
different data-base scales. In a particular case of just one fixed
scale, one could accomplish the normalization with respect to the
summation over $k$, which would define the probability weights for
a restricted data-base ensemble, as was postulated in reference
[37]. The derived scenario probability (44) concerns the general
case of an arbitrary ensemble $\{{\bf D}_k(j)\}$ of the data bases
(36). Being general, the approach of the present paper makes it
possible to answer several principal questions.

One important question concerns the choice of the data-base order.
More in detail, the problem is as follows. For a fixed time scale,
labelled by $j$, we may analyze different data bases ${\bf
D}_k(j)$, with varying $k=1,2,3,\ldots$. Then, how many terms
$f_k^{(j)}$ should we take? That is, when should we stop
increasing $k$? The answer is straightforward: The data-base order
$k$ has to be increased till we reach numerical convergence. More
precisely, this means the following. Let us be satisfied by the
results of an error $\varepsilon$. Then we need to increase $k$ up
to the {\it saturation number} $N_j = N_j(\varepsilon)$ such that
\begin{equation}
\label{45}
| f_{k+n}^*(j,t) - f_k^*(j,t)| < \varepsilon
\end{equation}
for $k\geq N_j$, all $n\geq 0$, and $t\in[0,1]$. For this $k=N_j$,
we have the {\it saturated data base}
${\bf D}(j)\equiv {\bf D}_{N_j}(j)$ and the related {\it
saturated forecast}
\begin{equation}
\label{46}
f^*(j,t) \equiv f_{N_j}^*(j,t)
\end{equation}
characterized by the {\it saturated scenario probability}
\begin{equation}
\label{47}
p(j,t) \equiv p_{N_j}(j,t) \; .
\end{equation}
Thus, varying the data-base scales, labelled by $j$, we get the
{\it saturated ensemble} $\{ f^*(j,t)\}$ of scenarios (46), with the
probability measure (47).

The {\it most probable scenario} from the ensemble $\{ f^*(j,t)\}$
is the forecast $f^*(j_0,t)$ having the largest probability, such
that
\begin{equation}
\label{48}
\max_j p(j,t) = p(j_0,t) \; .
\end{equation}
The latter, because of the form (44), is equivalent to the
condition of the minimum for the absolute value of the multiplier
$m(j,t)\equiv m_{N_j}(j,t)$, so that
\begin{equation}
\label{49}
\min_j|m(j,t)| =|m(j_0,t)| \; .
\end{equation}
Having the probability measure (47), it is possible to define the
average forecast, or the {\it expected forecast}
\begin{equation}
\label{50}
<f(t)>\; =\sum_j\; p(j,t)\; f^*(j,t) \; .
\end{equation}
The dispersion
$$
\sigma^2(t) \equiv \; <f^2(t)>\; - \; <f(t)>^2 \; ,
$$
in the case of a market, describes the market volatility. The
latter can also be characterized by the variance coefficient
$\kappa(t)\equiv\sigma(t)/<f(t)>$.

The last question to be answered in order to have a completely
self-consistent theory is how to choose the data-base time scale.
In general, the moments of time $t_n^{(j)}$ should not be
compulsory equidistant. However, for practical purpose, it looks
more convenient to take them as such, defining the time step as
$\Delta_j\equiv t_n^{(j)} - t_{n+1}^{(j)}$. It seems natural to
start with the time scale $\Delta_0=1$, which equals the
prediction horizon. Then, one may decrease as well as increase the
time step, for instance, according to the rule $\Delta_{2j}=
2^{-j},\; \Delta_{2j+1} = 2^j$, with $j=0,1,2,\ldots$. One has to
stop decreasing and increasing the data-base time scale at such
$j=j_{max}$, when numerical convergence is reached. This implies
that, for a given error $\varepsilon$, one gets the inequality
$$
| f^*(j+m,t) - f^*(j,t)| < \varepsilon
$$
for $j\geq j_{max}$, all $m\geq 0$, and $t\in[0,1]$.

\vskip 5mm

{\large{\bf 5. Summary of main ideas}}

\vskip 3mm

In this paper, a novel approach to analyzing time series has been
presented. There are several principal points distinguishing this
approach from the standard one. The aim of this paper has been to
clearly describe these principal points forming the body of a
self-consistent theory. Not yet all parts of this theory have been
exploited in full for practical applications; the calculational
work is in process. But some simplified versions of the approach
have been illustrated by a number of examples for market time
series [37,38]. In this concluding section, I would like to
emphasize again the main ideas the approach is based on.

To better stress the principal difference of the present approach
from the standard way of analyzing time series, let us recall the
basic idea of the latter: For a given set of data $\{ f_n|\;
n=0,1,2,\ldots,k\}$ one tries to invent a relation
$f_t=f(f_0,f_1,\ldots,f_k,t,\xi)$ connecting the value $f_t$ at
the moment of time $t$ with the past data. This relation can be in
the form of an explicit function or in the form of
a difference or differential equation, including a
stochastic term $\xi$ modelling noise [1--5]. There are, to my
mind, two principal deficiencies of such an approach. First, I
think that no explicit equations, indifferently to how elaborated
they are, can grasp all peculiarities of a realistic complex
system, such as a market. Second, any given relations reflect only
the past history, providing an {\it interpolation} for the
learning historical period, while for predicting future one needs
an {\it extrapolation}.

The self-similar approach is based not on attempts to invent a
relation between the historical points but it tries to discover
{\it dynamic trends} resulting in these points. For this purpose,
instead of studying relations between points, it is necessary to
analyze {\it relations between sequences}. Interpolative formulas
are used here only as a starting step. The comparison of different
interpolative expressions makes it possible to produce an
extrapolation, that is, forecasting.

The basic philosophical idea of the self-similar analysis of
complex systems is the {\it Law of Self-Similar Evolution},
formulated in Sec. 1. And the mathematical foundation is provided
by the self-similar extrapolation of asymptotic series, described
in Sec. 2. This extrapolation can be reformulated as forecasting
for time series, as is done in Sec. 3. Since any forecasting can
only be probabilistic, the way of defining the related probability
measure is explained in Sec. 4. Some details of the self-similar
analysis can be changed. For example, the number of points in a
given data base can be reduced by replacing several neighbouring
values $f_n$ by either arithmetic averages or by fitting the
values on a large time interval with the help of simple splines
[34]. However, in the process of these kinds of averaging, some
information on the considered time series will be inavoidably
lost.

Another change could concern the definition of the effective
control time $\tau_n$. The latter is to be defined from the
minimization of a cost functional. As an example, the
time-distance cost functional (21) was considered. But, in
general, one could opt for another cost functional, depending on
the available information and imposed constraints.

Also, it would be possible to deal not with the initially given
data base but with some transforms of it. For instance, one could
consider the set $\{\ln f_n\}$ instead of $\{ f_n\}$. Or one could
keep in mind a more elaborated transform, like the wavelet
transforms, often employed for analyzing time series [39].

It is worth stressing once more that the notion of self-similarity
exploited throughout the paper is understood here as the {\it group
self-similarity}, which is a more general notion than the trivial
geometric self-similarity one usually talks about in connection
with fractals. In the latter case one assumes the existence of the
scaling relation $f(\lambda x)=\lambda^\alpha f(x)$ for the
considered function $f(x)$. Such a relation, with a given
boundary, or initial, condition $f(x_0)=f_0$, immediately results
in the power-law function $f(x)=f_0(x/x_0)^\alpha$. One could
consider a slightly more complicated scaling relation as
$f(\lambda x)=u(\lambda)f(x)$, with a known function $u(\lambda)$.
However again, with the given boundary condition, this immediately
gives the answer $f(x)=f_0/u(x_0/x)$. All such scaling relations
produce the considered function in an explicit form. Whereas the
group self-similarity (13) provides an equation that is yet to be
solved.

An important point is that the group self-similarity employed
here, and which is the basis of the {\it Self-Similar
Approximation Theory} [13--23], has to do not with a scaling of a
variable but with the motion with respect to the effective time
whose role is played by the approximation number. Here it is the
{\it motion on the manifold of approximants}.

Finally, if we would like, digressing from mathematical
foundations, to briefly conclude why the idea of the group
self-similarity does work for extrapolating asymptotic series and
forecasting time series, then we should return back to the {\it
Law of Self-Similar Evolution} telling that all complex systems
develop self-similarly, preserving their generic features in the
course of their evolution. Such features may be not noticeable
from the first glance but rather hidden somewhere in genes. The
group self-similarity is a kind of {\it genetic self-similarity}.
Remember also that the Lord created the man in a self-similar way [40].

\vskip 5mm

{\bf Acknowledgenemts}

\vskip 3mm

I am grateful to M. Ausloos and E. Yukalova for the interest to my work
and useful discussions.

\end{document}